\documentclass[modern]{aastex631}
\usepackage{xspace,graphics,graphicx}

\newcommand\teff{\ensuremath{T_\mathrm{eff}}\xspace}
\newcommand\logg{\ensuremath{\log g}\xspace}
\newcommand\msun{\ensuremath{M_\odot}\xspace}
\newcommand\tcool{\ensuremath{\tau_\mathrm{cool}}\xspace}
\newcommand\tnuc{\ensuremath{\tau_\mathrm{nuc}}\xspace}

\accepted{July 12, 2021}
\submitjournal{The Astronomical Journal}

\shorttitle{The IFMR for non-DA WDs}
\shortauthors{Barnett et al.}
\graphicspath{{./}{figs/}}

\begin{document}

\title{The Initial-Final Mass Relation for Hydrogen-Deficient White Dwarfs \footnote{Some of the data presented herein were obtained at the W.M. Keck Observatory, which is operated as a scientific partnership among the California Institute of Technology, the University of California and the National Aeronautics and Space Administration. The Observatory was made possible by the generous financial support of the W.M. Keck Foundation.}}


\correspondingauthor{Kurtis  A. Williams}
\email{Kurtis.Williams@tamuc.edu,bedard@astro.umontreal.ca,bolte@ucolick.org}

\author{Joseph W. Barnett}
\affiliation{Texas A\&M University-Commerce  \\
P.O. Box 3011 \\
Commerce, TX 75429-3011, USA}

\author[0000-0002-1413-7679]{Kurtis A. Williams}
\affiliation{Texas A\&M University-Commerce  \\
P.O. Box 3011 \\
Commerce, TX 75429-3011, USA}

\author[0000-0002-2384-1326]{A.\ B\'{e}dard}
\affiliation{D\'{e}partement de Physique \\
Universit\'{e} de Montr\'{e}al \\ 
Montr\'{e}al, QC H3C 3J7, Canada}

\author{Michael Bolte}
\affiliation{University of California Santa Cruz \\ 
University of California \\
1156 High St. \\ 
Santa Cruz, CA 95064, USA}

\begin{abstract}
The initial-final mass relation (IFMR) represents the total mass lost by a star during the entirety of its evolution from the zero age main sequence to the white dwarf cooling track.  The semi-empirical IFMR is largely based on  observations of DA white dwarfs, the most common spectral type of white dwarf and the simplest atmosphere to model.  We present a first derivation of the semi-empirical IFMR for hydrogen deficient white dwarfs (non-DA) in open star clusters.  
We identify a possible discrepancy between the DA and non-DA IFMRs, with non-DA white dwarfs $\approx 0.07\,\msun$ less massive at a given initial mass.  Such a discrepancy is unexpected based on theoretical models of non-DA formation and observations of field white dwarf mass distributions. If real, the discrepancy is likely due to enhanced mass loss during the final thermal pulse and renewed post-AGB evolution of the star.  However, we are dubious that the mass discrepancy is physical and instead is due to the small sample size, to systematic issues in model atmospheres of non-DAs, and to the uncertain evolutionary history of Procyon B (spectral type DQZ). A significantly larger sample size is needed to test these assertions.  In addition, we also present Monte Carlo models of the correlated errors for DA and non-DA white dwarfs in the initial-final mass plane.  We find the uncertainties in initial-final mass determinations for individual white dwarfs can be significantly asymmetric, but the recovered functional form of the IFMR is grossly unaffected by the correlated errors.

\end{abstract}

\section{Introduction \label{sec.intro}}

White dwarfs (WDs) are the endpoint of stellar evolution for the vast majority of stars.  These stellar remnants thus provide an observational anchor for studies of the late stages of stellar evolution and serve as forensic evidence of the complex physical processes active in the cores of asymptotic giant branch (AGB) stars.

Most ($\sim 80\%$) WDs are in the DA spectral class \citep[e.g.,][]{1999ApJS..121....1M,2019MNRAS.486.2169K}, meaning that they are enshrouded by an opaque, nearly pure hydrogen layer with a mass of $\sim 10^{-10} - 10^{-4}\,M_\mathrm{WD}$.  These atmospheres are relatively straightforward to model at warmer temperatures, though recent progress in computational and laboratory physics has revealed systematic issues related to molecular interactions and line profiles that still are not fully understood \citep[e.g.,][]{2019ApJ...885...86S,2020PhRvL.124e5003G}. Nonetheless, observations of DA WDs can be matched to model atmospheres to determine fundamental parameters such as the effective temperature,  \teff,  and surface gravity, \logg, to relatively high precision \citep[e.g.,][]{2005ApJS..156...47L,2009A&A...505..441K}.

The non-DA spectral classes are generally hydrogen deficient; most have atmospheres dominated by helium, though trace residuals of hydrogen or metals are often present \citep[e.g.,][]{2019ApJ...876...67B,2019ApJ...882..106G,2019ApJ...885...74C}.  Most non-DA WDs also are suspected to change spectral type over their evolution due to atmospheric chemical stratification from to the WD's high surface gravity and the establishment and growth of a surface convective zone; the exact evolution will depend on the mass of residual hydrogen (if any) and the thickness of the He layer \citep[e.g.,][]{2018ApJ...857...56R,2019ApJ...878...63B,2019ApJ...885...74C,2020MNRAS.492.3540C,2020ApJ...901...93B}.  

A primary channel for non-DA WD production is via a late thermal pulse (LTP), a final burst of nuclear fusion that occurs during post-AGB evolution, or a very late thermal pulse (VLTP), a thermal pulse that occurs after the star has completed post-AGB evolution and entered the WD cooling track \citep{1979A&A....79..108S,1982ApJ...263L..23I,1983ApJ...264..605I}.  These pulses consume or drive off nearly all remaining hydrogen and much of the remaining He shell \citep[e.g.,][]{2005A&A...435..631A,2006PASP..118..183W}.

Only a few LTPs and VLTPs have been directly observed: \object{V4334 Sgr} \citep[Sakurai's Object; e.g.,][]{1996ApJ...468L.111D,2020MNRAS.493.1277E} and \object{V605 Aql}  \citep[e.g.,][]{1997AJ....114.2679C,2006ApJ...646L..69C} are both suspected VLTPs \citep[c.f.][]{2011MNRAS.410.1870L,2013ApJ...771..130C}, while \object{FG Sge} \citep[e.g.,][]{1998ApJS..114..133G,2003ApJ...583..913L} and \object{V839 Aql} \citep[e.g.,][]{1993A&A...267L..19P,2014A&A...565A..40R,2017MNRAS.464L..51R} are classified as likely LTPs.  These thermal pulses do drive some additional mass loss; the ejected gas mass is measured at $\sim 10^{-5} - 10^{-4} \,\msun$ based on observations of ejecta in known (V)LTP events and in H-deficient planetary nebulae \citep[e.g.,][]{2013ApJ...771..130C,2018NatAs...2..784G,2020MNRAS.493.1277E,2020arXiv201204223S}.  These thermal pulses send the star back toward the AGB in the Hertzsprung-Russell diagram, and total mass loss in the renewed post-AGB phase may approach to $\sim 10^{-2}\,\msun$ \citep{2005A&A...435..631A}.  This additional mass loss would therefore only be a small fraction of the overall WD mass.

\subsection{The Initial-Final Mass Relation}

The initial-final mass relation (IFMR) quantifies the correspondence between the zero-age main sequence mass of a star and the mass of its WD progeny.  The IFMR is generally assumed to be single-valued with negligible intrinsic scatter, weakly dependent on progenitor metallicity, and applicable only in the case of single star evolution or components of wide binary systems sufficiently separated that stellar interactions have no appreciable impact on each component's evolution.

Theoretically, the IFMR can be calculated from evolutionary models of late stage AGB stars.  This model IFMR takes the WD mass to be the core mass of a star at the  cessation of primary nuclear burning stages; post-AGB mass loss and residual nuclear burning should have little effect on the remnant mass.  In reality, physical limitations on models of processes such as core erosion during third-dredge up, convective and rotational mixing, and core growth during an uncertain number of successive thermal pulses limit the accuracy of these models \citep[e.g.,][]{2019ApJ...871L..18C}.  For these reasons, many published modeled IFMRs tend to relate the progenitor star mass to the core mass at the first thermal pulse \citep[e.g.,][]{2013MNRAS.434..488M}.

The IFMR can also be derived semi-empirically through the observations of WDs in simple stellar systems such as open star clusters or wide binary systems in the field.  In these cases the total system age is determined to reasonable accuracy through methods such as isochrone fitting (in an open cluster setting), and the cooling age of the white dwarf, \tcool, can be determined by measuring the WD mass and effective temperature and then interpolating WD evolutionary model grids.  The difference between the cluster age and \tcool gives the nuclear lifetime of the progenitor star, \tnuc, which corresponds to the progenitor star's mass.  This strong reliance on stellar and WD evolutionary models is why this observationally-based IFMR is known as ``semi-empirical," and a number of the most important systematics affecting the semi-empirical IFMR are quantified in \citet{Marigo2008} and still remain unresolved. Nonetheless, the semi-empirical IFMR has converged in recent years to a relatively tight relation with little intrinsic scatter \citep[e.g.,][]{2018ApJ...866...21C, 2018ApJ...867...62W}, at least for WDs arising from nearly solar metallicity progenitors that are analyzed using consistent methods with primarily self-consistent models.

Published semi-empirical IFMRs are almost entirely constructed from WDs of spectral type DA.  For the $\approx 80\%$ of WDs that are DAs, this is not problematic.  However, the non-DA WDs have undergone additional evolutionary processes that have resulted in the loss of most if not all of their outer H layer.  It therefore is appropriate to consider the possibility that the IFMR for non-DA WDs may differ from the DA IFMR.

A priori one would expect little if any difference in the DA and non-DA IFMRs.  The (V)LTP channel for production of non-DA WDs does not appear to result in significant mass loss from the evolving stellar core \citep[e.g.,][]{2013ApJ...771..130C,2018NatAs...2..784G,2020MNRAS.493.1277E,2020arXiv201204223S}.  Empirically, the mean masses and overall mass distribution of field DA and non-DA WDs are observed to be very similar \citep[spectroscopic mean masses of $0.615\,\msun$ for DAs and $0.625\,\msun$ for DBs;][]{2019ApJ...871..169G}, though the DA distribution has a more substantial high-mass tail and a secondary peak at low mass due to binary star evolution \citep[e.g.,][]{Bergeron1992,2010ApJ...712..585F,2012ApJ...757..116F,2013ApJS..204....5K,2019ApJ...882..106G}.

No substantial attempt has been made at constructing the semi-empirical, non-DA IFMR to date.  Non-DA WDs are rarer than DA WDs, so small number statistics will impact the results.  More importantly, though, is the inherent difficulty in determining precise and accurate parameters for non-DA WDs \citep[e.g.,][]{2019MNRAS.482.5222T,2019ApJ...876...67B,2019ApJ...882..106G,2020ApJ...901...93B}.  Processes such as van der Waals broadening, non-ideal atomic effects, and convective energy transport remain significant sources of uncertainty \citep[e.g.,][]{2011ApJ...737...28B,2015A&A...583A..86K,2018MNRAS.481.1522C}.    Although substantial progress is being made through more complex computational efforts \citep{2020ApJ...901..104T,2021MNRAS.501.5274C} and even laboratory studies of He at WD temperatures and pressures \citep{2020IAUS..350..231M}, observational determinations of non-DA WD parameters are simply subject to larger systematic uncertainty than those of DA WDs.

In this paper we present an attempt to construct the semi-empirical IFMR for non-DA WDs and compare that to both the DA semi-empirical IFMR and model IFMRs.  We discuss the construction of a suitable non-DA sample from both previously-published and new observational data for open cluster non-DA WDs in Section \ref{sec.obs}.  We then discuss new modeling of the observational uncertainties in the semi-empirical IFMR that could affect the interpretation of both the DA and non-DA IFMRs in Section \ref{sec.errors}.  Finally, we discuss apparent differences between the DA and non-DA IFMRs that point out the need for additional study in Section \ref{sec.db_ifmr}.

\section{Observations \& Analysis \label{sec.obs}}

\subsection{Sample Selection \label{sec.sample}}
For this work, we scoured the literature for non-DA WDs in the fields of open star clusters.  This search uncovered 13 non-DA WDs: two in the field of M34 \citep{2008AJ....135.2163R}, one in the Hyades \citep[e.g.,][]{1976ApJ...210..524G,1984AJ.....89..830E,2012AAp...547A..99T}, one in the field of M47 \citep{2019ApJ...880...75R}, two in the field of M35 \citep{2006ApJ...643L.127W,2009ApJ...693..355W}, five in the field of M67 \citep{2018ApJ...867...62W}, and two in the field of \object{NGC 6633} \citep{Williams2007}. We also include Procyon B, as its mass, effective temperature, and the total system age are tightly constrained \citep{2015ApJ...813..106B,2018RNAAS...2..147B,2019MNRAS.482..895S}, allowing us to consider the binary as a two-star cluster.  

We then searched the Gaia EDR3 catalog \citep{2020arXiv201201533G} for each non-DA WD; seven WDs have tabulated proper motion and parallax measurements with the renormalized unit weight error \citep[RUWE,][]{2020arXiv201206242F,2020arXiv201203380L} $\leq 1.4$.  We consider a WD to be a member of its respective open star cluster if the proper motion vectors and measured parallax are within $2\sigma$ of the cluster measurements tabulated in \citet{2018AAp...616A..10G}.  These seven WDs, their Gaia EDR3 astrometry, and cluster membership determinations are presented in Table~\ref{tab.astrometry}. 

Due to their faint apparent magnitudes, four non-DA WDs in M67 do not have astrometry in the Gaia EDR3 catalog but are proper motion members of the cluster \citep{2010AAp...513A..50B,2018ApJ...867...62W}: \object[WCB2018 WD18]{M67:WD18}, \object[WCB2018 WD27]{M67:WD27}, \object[LB 6373]{M67:WD30},
and \object[WCB2018 WD31]{M67:WD31}.  

We exclude two non-DAs in the field of the open cluster Messier 35 due to a lack of proper motion measurements:  \object[LAWDS NGC 2168 4]{M35:LAWDS~4}, a DB WD, and \object[WD 0605+243]{M35:LAWDS~28}, a hot DQ WD.  Attempts to measure the proper motions of these two WDs are underway, and so we will wait to analyze these objects until the results are known.

Our final sample of cluster member non-DA WDs is given in Table~\ref{tab.specfits}.

\begin{deluxetable*}{lcCCCCCCc}
\tabletypesize{\scriptsize}
\tablecolumns{9}
\tablewidth{0pt}
\tablecaption{Astrometric Cluster Memberships of Non-DA WDs in Gaia EDR3 \label{tab.astrometry}}
\tablehead{ \colhead{WD ID} & \colhead{Gaia EDR3} & \dcolhead{\mu_\alpha} & \dcolhead{d\mu_\alpha} & \dcolhead{\mu_\delta} & \dcolhead{d\mu_\delta} & \dcolhead{\varpi} & \dcolhead{d\varpi} & \colhead{Cluster} \\
  & Source ID &  \mathrm{mas\ yr}^{-1} & \mathrm{mas\ yr}^{-1} & \mathrm{mas\  yr}^{-1} & \mathrm{mas\ yr}^{-1} & \mathrm{mas} & \mathrm{mas} & Member?  }
\startdata
\object[WD 0237+425]{M34:LAWDS 7} & 337164523400019456 & 40.93 & 0.37 & -18.37 & 0.35 & 5.85 & 0.36 & No \\
\object[WD 0238+427.2]{M34:LAWDS 26} & 337191594578690560 & 9.0 & 1.5 & -6.0 & 1.2 & -0.9 & 1.1 & No \\
\object[EGGR 316]{Hyades:EGGR 316} & 3308403897837092992 & 95.286 & 0.039 & -20.835 & 0.026 & 21.760 &  0.033 & Yes \\ 
\object[Gaia DR2 3029912407273360512]{M47:DBH} & 3029912407273360512 & -8.10 & 0.41 & 2.11 & 
0.36 & 2.53 & 0.50 & Yes \\
\object[LB 3600]{M67:WD21} & 604898490980773888 &  -11.4 & 0.43 & -3.43 & 0.32 & 0.98 & 0.44 & Yes \\
\object[WD 1824+063]{NGC6633:LAWDS 14} & 4477166581862730752 & 13.584 & 0.074 & -78.115 & 0.071 & 10.709 & 0.067 & No \\
\object[WD 1825+067]{NGC6633:LAWDS 16} & 4477253202776118016 & -5.32 & 0.64 & -9.62 & 0.60 & -1.04 & 0.61 & No 
\enddata
\end{deluxetable*}

\begin{deluxetable}{lcCCCCl}
\tablewidth{0pt}
\tablecaption{Non-DA WD Sample and Adopted Parameters \label{tab.specfits}}
\tablehead{\colhead{WD} & Spectral & \teff & \logg & M_\mathrm{WD} & \log \tau_\mathrm{cool} & Reference \\
& Type & \mathrm{K} & \mathrm{cgs} & \msun & \log \mathrm{yr} & }
\startdata
EGGR 316 & DBA & 15,129\pm 361 & 8.25\pm 0.07 & 0.74\pm 0.06 & 8.49\pm 0.07 & 1 \\
M47:DBH & DBH & 32,100\pm 2700 & 8.73\pm 0.09 & 1.06\pm 0.05 & 7.84\pm 0.20 & 2 \\
Procyon B & DQZ & 7740\pm 50 & 8.02\pm 0.02 & 0.59\pm 0.01 & 9.12\pm 0.01 & 3,4 \\
M67:WD18 & DBA & 11,820\pm 1313 & 8.04\pm 0.24 & 0.61\pm 0.14 & 8.65\pm 0.21 & This work \\
M67:WD21 & DB & 30,765\pm 2757 & 7.84\pm 0.10 & 0.53\pm 0.05 & 6.93\pm 0.14 & This work \\
M67:WD27 & DC & 7725\pm 520 & 7.65\pm 0.24 & 0.39\pm 0.11 & 8.91\pm 0.14 & This work \\
M67:WD30 & DB & 19,892\pm 477 & 8.19\pm 0.15 & 0.71\pm 0.09 & 8.04\pm 0.16 & This work \\
M67:WD31 & DC & 8732\pm 777 & 7.22\pm 0.24 & 0.23\pm 0.11 & 8.58\pm 0.15 & This work \\
\enddata
\tablerefs{ (1) \citet{2018ApJ...857...56R}, (2) \citet{2019ApJ...880...75R}, (3) \citet{2015ApJ...813..106B},  (4) \citet{2018RNAAS...2..147B} }
\end{deluxetable}

\subsection{White Dwarf Parameter Determination \label{sec.params}}

\subsubsection{Spectroscopic Observations \& Reduction \label{sec.specobs}}

We now discuss the spectral observations for the M67 non-DA WDs without previous parameter determinations.
The $3500-5500$ \AA\  observations for the cluster sample WDs have already been detailed in \citet{2009ApJ...693..355W} and \citet{2018ApJ...867...62W}.  In summary, spectra were obtained over multiple observing runs using the blue channel of the Low Resolution Imaging Spectrometer (LRIS) on the Keck I telescope \citep{1998SPIE.3355...81M} and extracted using the onedspec and twodspec packages in IRAF \citep{1986SPIE..627..733T,1993ASPC...52..173T}.  The extracted and co-added spectra have spectral resolutions of $\approx 8$~\AA~ FWHM.  

WDs with helium-dominated atmospheres are often mixed with small amounts of hydrogen; the H/He ratio is important to quantify, as it can significantly affect the measured \teff and \logg.  Due to pressure effects, often only the H$\alpha$ Balmer line is clearly present in optical spectra.  

In our previous publications on cluster WDs, the red channel spectroscopy covering the H$\alpha$ line were ignored, as H$\beta$ and higher order Balmer lines are sufficient to constrain \teff and \logg in DA WDs.  We therefore describe the red-channel LRIS spectral observations in more detail here; these data are used to constrain the atmospheric H/He ratio in our sample.  

The red-channel data were obtained simultaneously with the blue-channel data for each WD.  For all WDs, the D560 dichroic was used in combination with the 600 lines mm$^{-1}$,  7500~\AA\  blaze grating.  The central wavelength was set to 6245~\AA,  resulting in spectral coverage from the dichroic cutoff through $\sim 7500$~\AA. Flat-fielding used the internal halogen lamps, and wavelength calibration was performed using arclamps of Ne, Hg and Zn. For all observations, either a 1\arcsec-wide longslit or a 1\arcsec-wide slitlet was used, resulting in a spectral resolution of 4.7~\AA\  FWHM as measured from night sky emission lines.

\begin{figure*}
    \centering
    \includegraphics[clip, trim=0.5cm 5.5cm 1.5cm 3.5cm, width=\columnwidth]{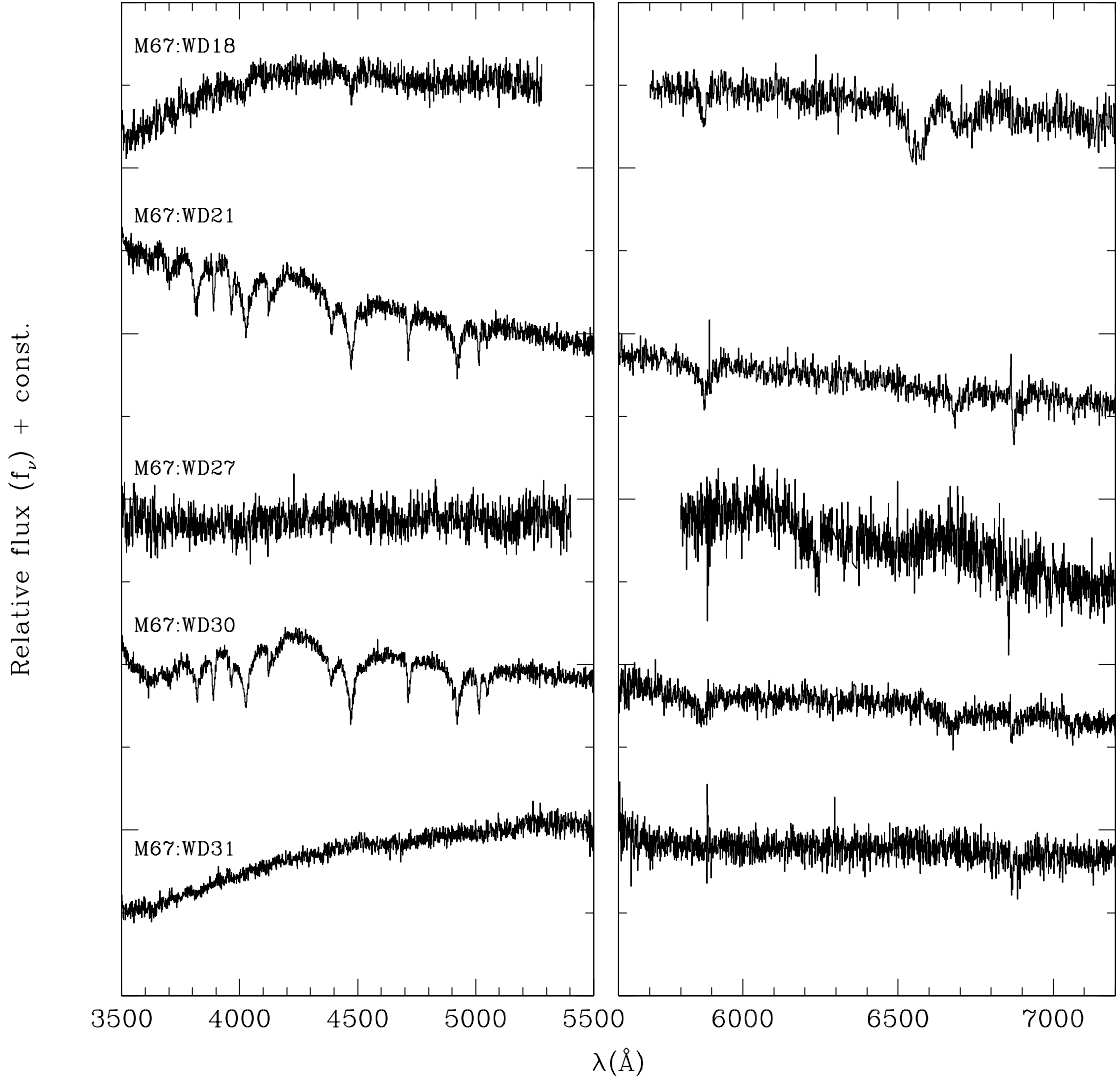}
    \caption{Spectra for the 5 non-DA WDs with new parameter determinations presented in this paper.  Spectra are relatively flux calibrated and offset vertically by an arbitrary constant.  \emph{Left panel:} Blue-side spectra, previously published in \citet{2018ApJ...867...62W}.  \emph{Right panel:} Red-side spectra, all newly presented here.  }
    \label{fig.spectra}
\end{figure*}

The data were reduced and spectra extracted using IRAF.  Multiple exposures were averaged using cosmic ray rejection and the spectra extracted using the IRAF twodspec package.  Relative flux calibration was performed using spectrophotometric standards observed the same night with the same instrumental configuration.  Details on variable observational parameters for each WD are given in Table~\ref{tab.obsparams}.  The spectra are plotted in Figure \ref{fig.spectra}.

\begin{deluxetable*}{lcccccc}
\tablewidth{0pt}
\tablecaption{Dates and Instrumental Setup for Spectral Observations \label{tab.obsparams}}
\tablehead{\colhead{WD} & \colhead{Obs.~Date} & \colhead{Exp.~Time} & \colhead{Aperture} &  \colhead{Spectrophotometric} & \colhead{Detector} & \colhead{S/N} \\
& (UT) & (s) & & Standard Applied & & Ratio\tablenotemark{a}}
\startdata
M67:WD18 & 2007 Jan 20 & $8\times 1200$ & 1\arcsec\ slitmask & \object{Feige 34} & Tektronix\tablenotemark{b} & 22 \\
M67:WD21 & 2007 Jan 20 & $2\times 600$ & 1\arcsec\ longslit & \object{Feige 34} & Tektronix\tablenotemark{b} & 32 \\
M67:WD27 & 2010 Feb 9 & $6\times 1140$ & 1\arcsec\ slitmask & \object{G 191-B2B} & LBNL\tablenotemark{c} & 22 \\
M67:WD30 & 2010 Feb 9 & $6\times 1140$ & 1\arcsec\ slitmask & \object{G 191-B2B} & LBNL\tablenotemark{c} & 32 \\
M67:WD31 & 2010 Feb 9 & $9\times 1140$ & 1\arcsec\ slitmask & \object{G 191-B2B} & LBNL\tablenotemark{c} & 21 \\
\enddata
\tablenotetext{a}{per resolution element, measured at 6200 \AA.}
\tablenotetext{b}{See \citet{1995PASP..107..375O} for details.}
\tablenotetext{c}{See \citet{2010SPIE.7735E..0RR} for details.}

\end{deluxetable*}

\subsubsection{Spectroscopic and Photometric Model Fitting}

In this section, we estimate the atmospheric parameters $\teff$ and $\logg$ of the M67 non-DA WDs by comparing photometric and spectroscopic data to the predictions of model atmospheres.

First, we analyze the photometric energy distribution of all five objects following the method described in \citet{2019ApJ...876...67B}. Briefly, observed magnitudes are dereddened and converted into average fluxes, which are then fitted to a grid of theoretical fluxes using the least-square Levenberg-Marquardt algorithm. We rely on the model atmospheres of \citet{2019ApJ...882..106G} and \citet{2019ApJ...878...63B} for DB(A) and DC stars, respectively. We fit $ugr$ photometry from \citet{2018ApJ...867...62W} for M67:WD18, M67:WD27, and M67:WD30, and $ugriz$ photometry from the Sloan Digital Sky Survey Data Release 16 \citep{2020ApJS..249....3A} for M67:WD21 and M67:WD31. We assume a distance of $883.0 \pm 0.9$ pc, obtained from the Gaia-derived cluster parallax $\varpi = 1.1325 \pm 0.0011$ mas \citep{2018AAp...616A...1G}, and use the dereddening procedure of \citet{2006AJ....131..571H}. Four out of five objects do not show H$\alpha$; in these cases, we simply suppose a pure He composition. This assumption is entirely appropriate for DB WDs, as adding a trace of H at the detection limit has no effect on the measured parameters (see \citealt{2019ApJ...871..169G}). However, this is not quite true for DC WDs: the derived mass can be very sensitive to the assumed H abundance (see \citealt{2019ApJ...876...67B}). Unfortunately, the uncertainty associated with the unknown H content of DC stars is unavoidable. For M67:WD18, which shows H$\alpha$, we experimented with various H abundances and we adopt $\log \mathrm{(H/He)} = -5$ (see below). Our photometric fits are displayed in Figure \ref{fig.phot_fits}.

\begin{figure*}
\centering
\begin{minipage}{0.49\columnwidth}
\includegraphics[clip, trim=2.5cm 3.5cm 5.5cm 5cm, width=\columnwidth]{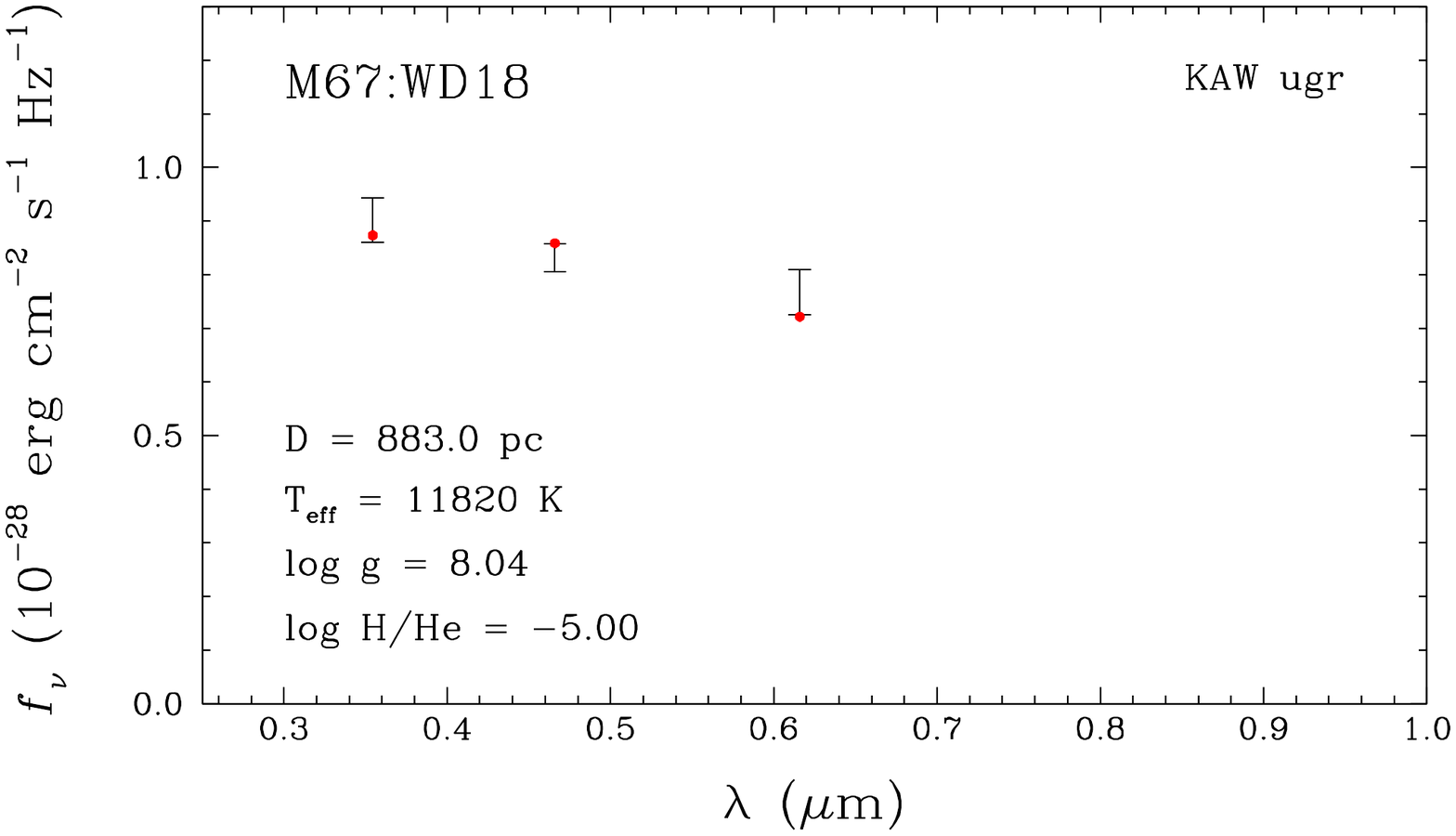}
\end{minipage}%
\begin{minipage}{0.49\columnwidth}
\includegraphics[clip, trim=2.5cm 3.5cm 5.5cm 5cm, width=\columnwidth]{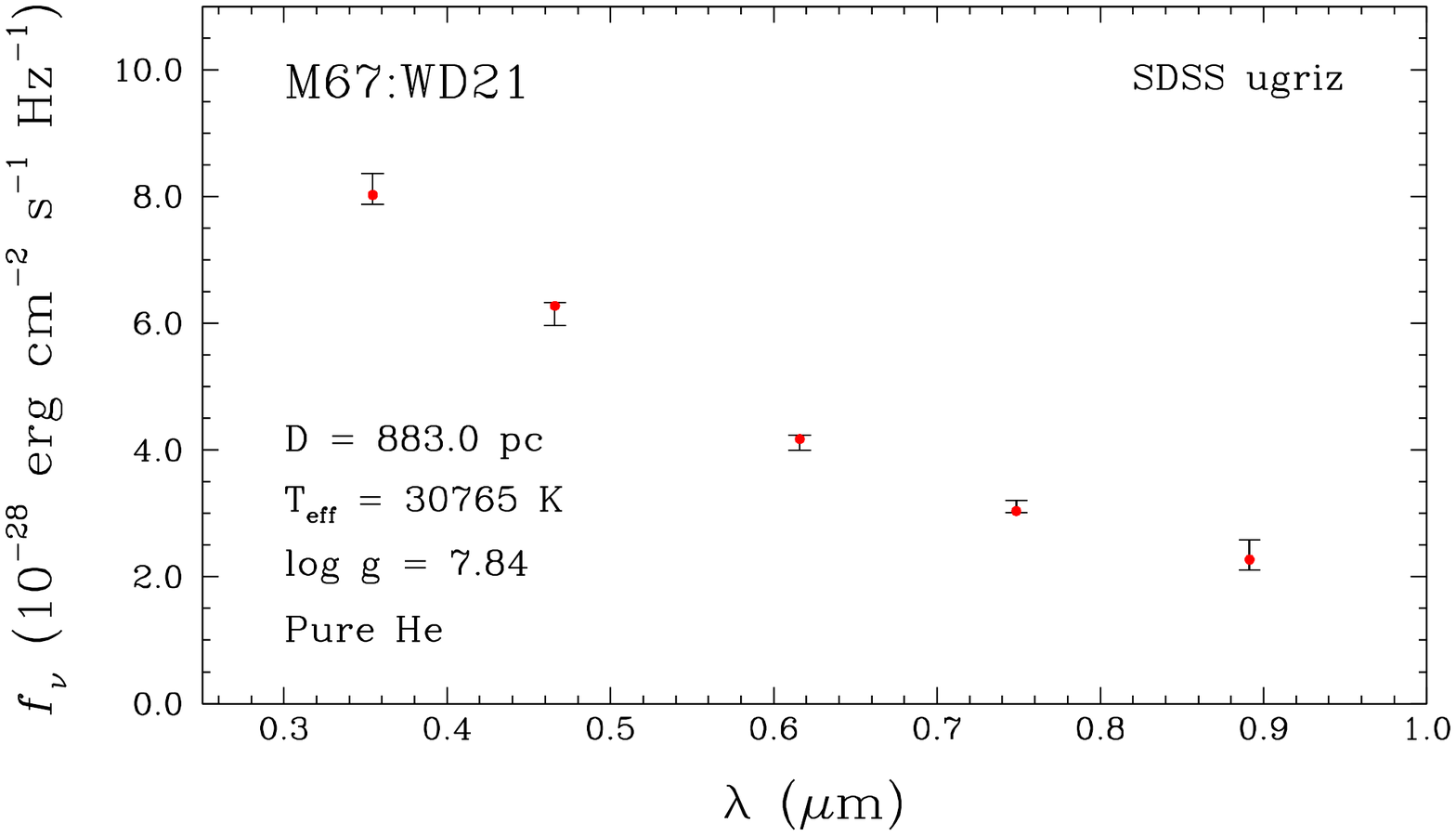}
\end{minipage}
\begin{minipage}{0.49\columnwidth}
\includegraphics[clip, trim=2.5cm 3.5cm 5.5cm 5cm, width=\columnwidth]{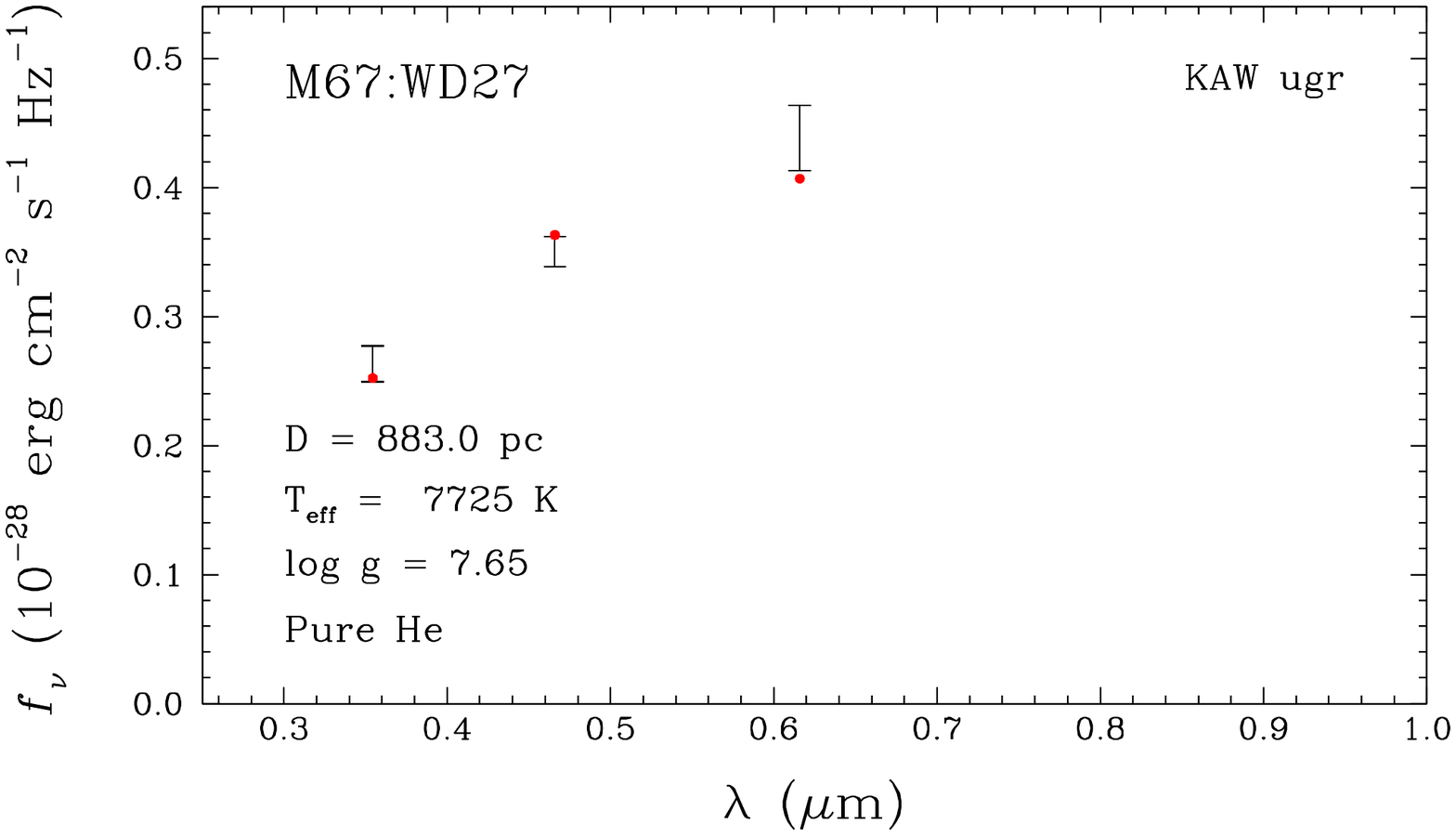}
\end{minipage}%
\begin{minipage}{0.49\columnwidth}
\includegraphics[clip, trim=2.5cm 3.5cm 5.5cm 5cm, width=\columnwidth]{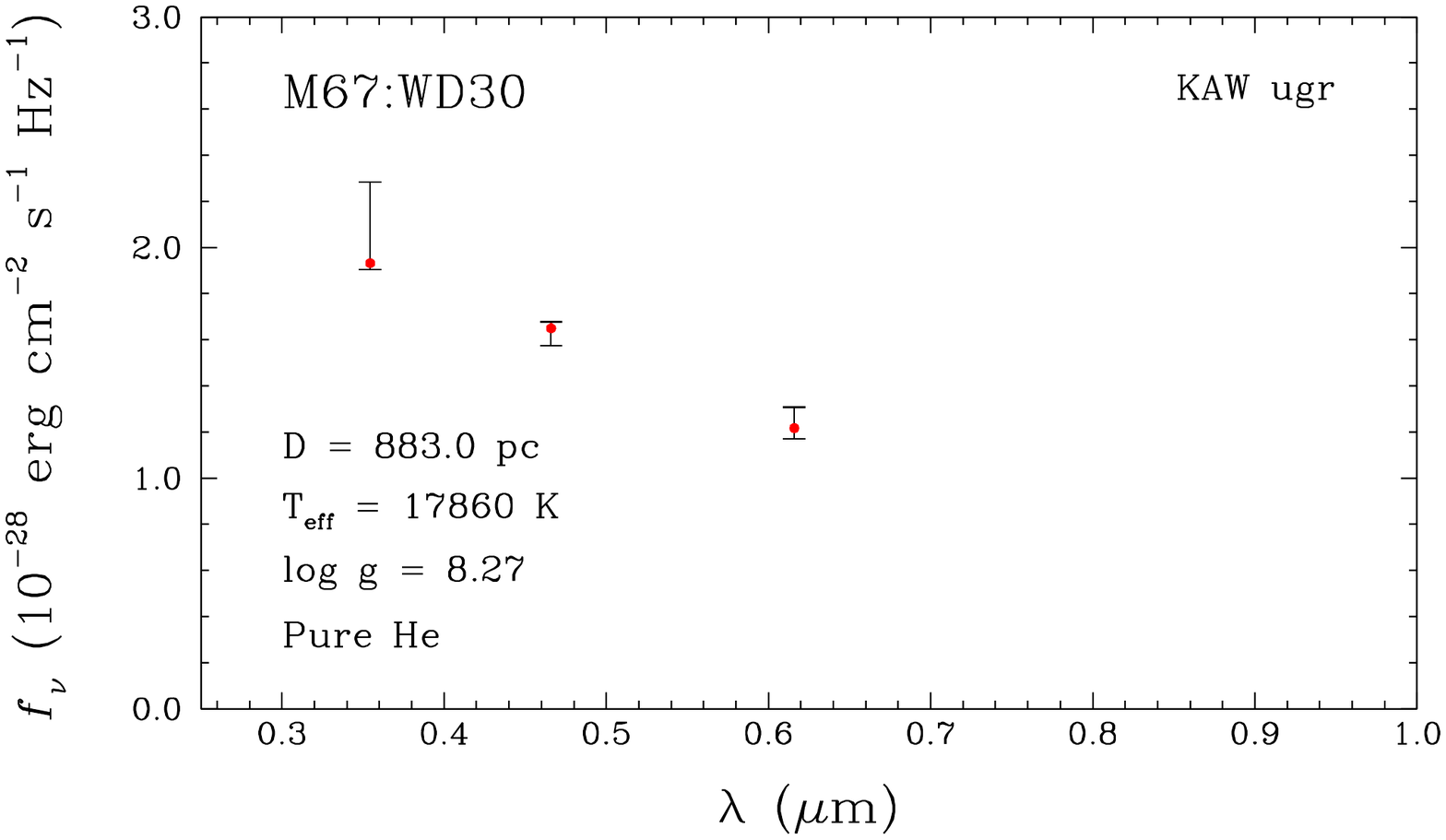}
\end{minipage}
\begin{minipage}{0.49\columnwidth}
\includegraphics[clip, trim=2.5cm 3.5cm 5.5cm 5cm, width=\columnwidth]{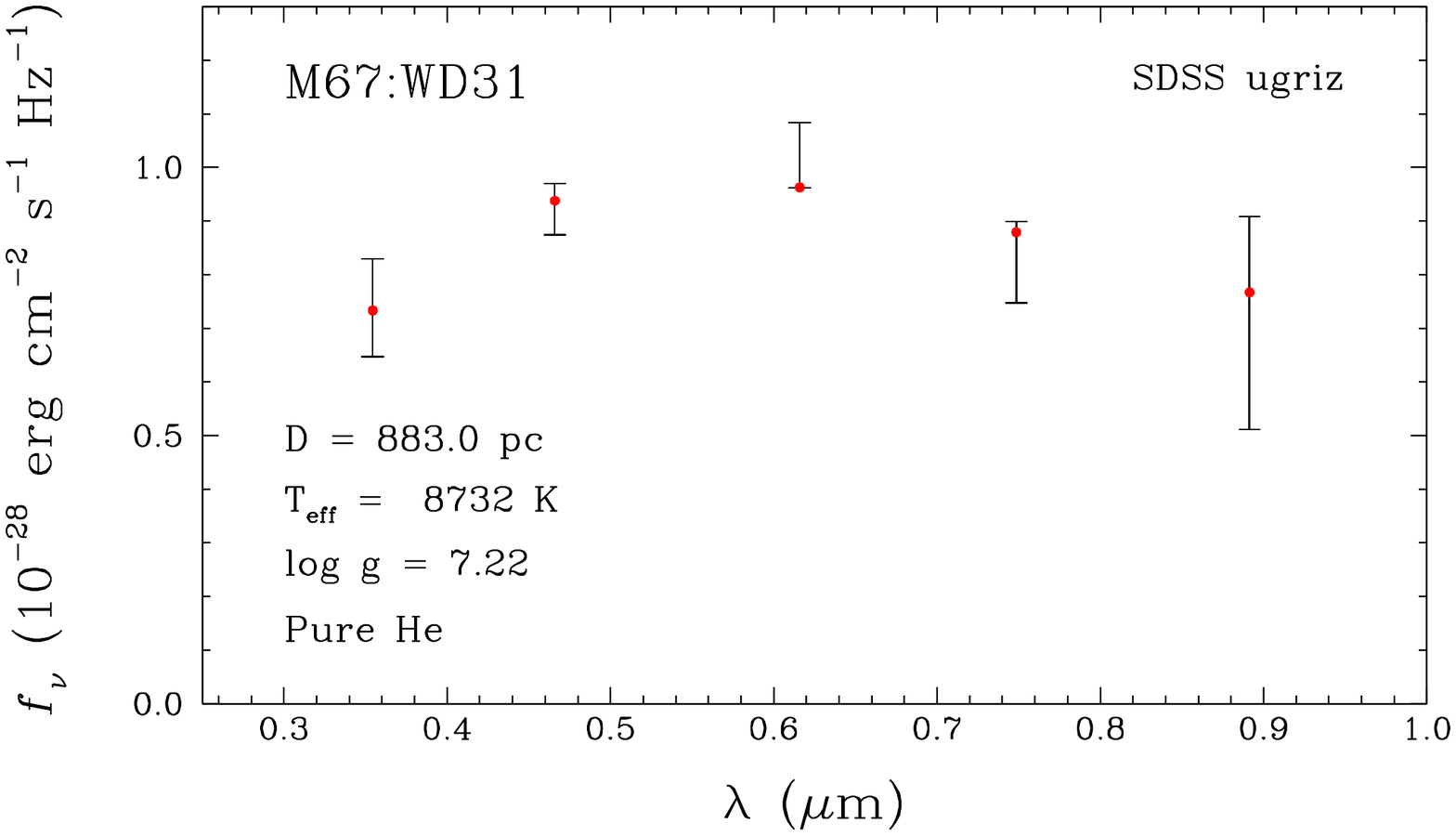}
\end{minipage}%
\begin{minipage}{0.49\columnwidth}
\hfill
\end{minipage}
\caption{Fits of observed photometry (error bars) to fluxes derived from synthetic DB spectra (red points).  The WD is identified in the upper left corner of each panel, and the best-fit solutions are given in the lower left.  In the upper right, the source of the observed photometry is indicated; ``KAW ugr" refers to $ugr$ photometry from \citet{2018ApJ...867...62W}, while ``SDSS ugriz" refers to PSF $ugriz$ photometry from the Sloan Digital Sky Survey Data Release 16 \citep{2020ApJS..249....3A}.}
\label{fig.phot_fits}
\end{figure*}

\begin{figure*}[tbp]
    \centering
    \includegraphics[clip, trim=0.5cm 5.5cm 1.5cm 9cm, width=\columnwidth] {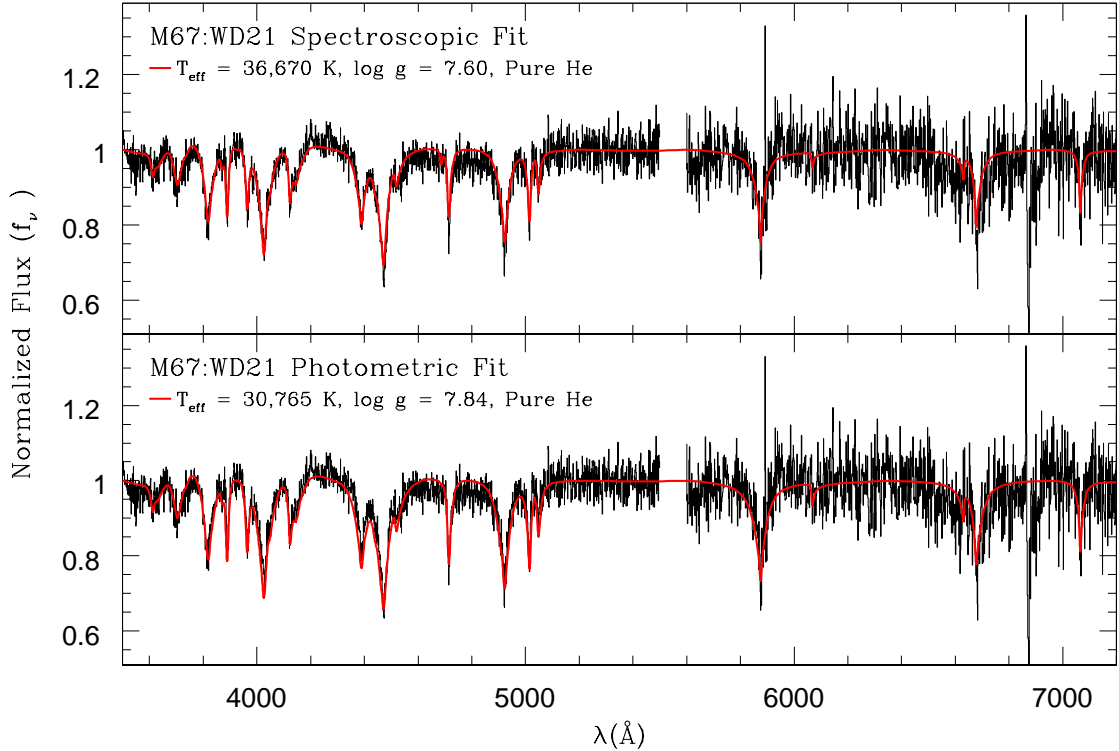}
    \caption{Comparison of spectral observations of M67:WD21 with model spectra.  \emph{(Top):} The normalized, observed spectrum (black) is compared to the best-fitting spectroscopic model (red).  Most of the \ion{He}{1} line core strengths are underpredicted.  \emph{(Bottom)}: The normalized observed spectrum with the synthetic spectrum for the best-fit photometric model overlaid.  Qualitatively the line cores appear to be better matched, but the strength of gravity-sensitive lines is overpredicted.}
    \label{fig.lb3600}
\end{figure*}

\begin{figure*}[tbp]
    \centering
    \includegraphics[clip, trim=0.5cm 5.5cm 1.5cm 13cm, width=\columnwidth]{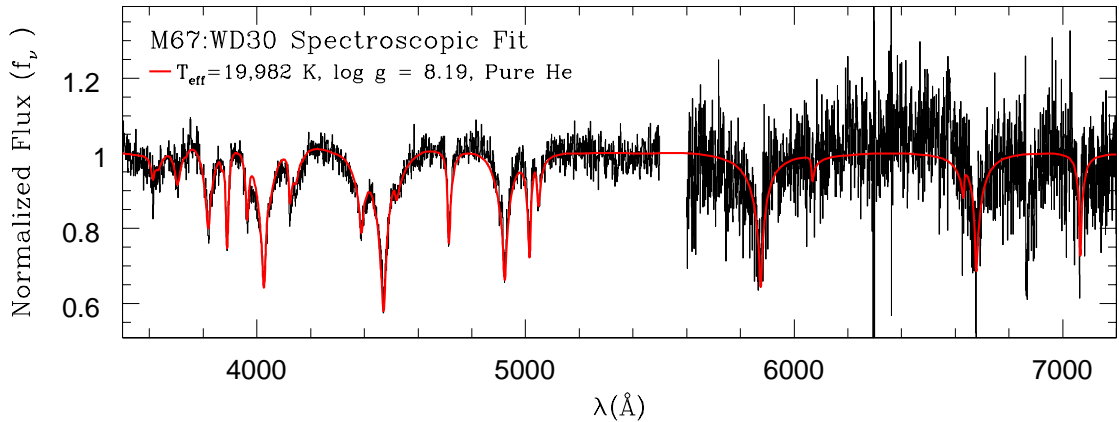}
    \caption{The normalized observed spectrum of M67:WD30 (black) compared to the best-fit normalized spectroscopic model (red).  Only the blue spectrum was used in fitting.  The dip in the red spectrum starting at 6861 \AA~ is the telluric B band.  Qualitatively the fit appears excellent, even on the red side. }
    \label{fig.wd30_spec}
\end{figure*}

Only two WDs (M67:WD21 and M67:WD30) exhibit \ion{He}{1} lines strong enough to allow a meaningful spectroscopic analysis. For these objects, we also derive $\teff$ and $\logg$ values based on the spectroscopic fitting method of \citet{2011ApJ...737...28B}. Briefly, the observed spectrum is first normalized to a continuum set to unity and then fitted to a grid of theoretical spectra using the least-square Levenberg-Marquardt algorithm. We employ the same model atmospheres mentioned above and again assume a pure He composition. We only include the blue spectrum, where most \ion{He}{1} lines are located, in the fitting process. Unlike photometric fits which directly yield reliable uncertainties in $\teff$ and $\logg$, spectroscopic fits only provide formal fitting errors which measure the ability of the models to reproduce the data and are often smaller than the true uncertainties in estimating the physical parameters. A more realistic estimate of the uncertainties can be obtained by analyzing multiple spectra of the same star and computing the standard deviation of the resulting parameters. Since we have only one spectrum for each object, we conservatively adopt the average spectroscopic uncertainties of the DB sample of \citet{2019ApJ...882..106G}: $\sigma_{\teff} = 0.024 \times \teff$ and $\sigma_{\logg} = 0.15$ dex. Our spectroscopic fits for M67:WD21 and M67:WD30 are shown in Figures \ref{fig.lb3600} and \ref{fig.wd30_spec}, respectively.

From each set of $\teff$ and $\logg$, we compute the corresponding stellar mass and cooling age using WD evolutionary sequences. For the DB(A) stars M67:WD18, M67:WD21, and M67:WD30, we rely on the C/O core, thin H envelope models of \citet{2020ApJ...901...93B}. For the DC stars M67:WD27 and M67:WD31, which have unusually low $\logg$ values, we use instead the He core models of \citet{2013A&A...557A..19A}. The results of our analysis are presented in Table 2. For the two objects with both spectroscopic and photometric parameters, we give the adopted solution (see below).

\subsubsection{Notes on Individual Objects \label{sec.notes}}

\paragraph{M67:WD21}  This hot, bright WD, also known as \object{LB 3600}, was first identified as a DB by \citet{1997ASSL..214...91F}.  Spectroscopic fitting provided two potential solutions, one on each side of the maximum strength of the \ion{He}{1} lines, but the hot solution is clearly preferred by the photometry.  Comparison of the observed spectrum to the best-fitting model shows a poor fit to the cores of most absorption lines (Figure \ref{fig.lb3600}, top panel).  This WD has SDSS photometry, and comparison of the observed spectrum to a synthetic spectrum with the photometric parameters shows a closer though still imperfect match to the line cores (Figure \ref{fig.lb3600}, bottom panel).  However, the photometric model does a poorer job of matching the gravity-sensitive 4388~\AA\  \ion{He}{1} line.

For this work, we adopt the photometric solution, in part because of the better match to the line cores, and in part because the larger error bars of the photometric fit more accurately define the uncertainty around the parameters for this WD.  A higher signal-to-noise spectrum would be useful for this DB.

\paragraph{M67:WD30} This DB gives the most satisfying spectral fit of the entire sample, as shown in Figure \ref{fig.wd30_spec}.  Although the spectral fit was determined only by the blue side spectrum, the synthetic spectrum for the red side qualitatively agrees with the observations very well.  The photometric parameters (Figure \ref{fig.phot_fits}) are entirely consistent with the spectroscopic parameters within the uncertainties.  In the following, we adopt the spectroscopic solution.  The derived WD mass of $0.71\pm 0.09\, \msun$ is larger than the masses of most M67 WDs; this WD could therefore be the remnant of blue straggler evolution \citep[see discussion in][]{2018ApJ...867...62W}.  

\paragraph{M67:WD18}  We classify this WD as a DBA due to the presence of a very strong H$\alpha$ absorption line.  Because this WD exhibits relatively weak \ion{He}{1} lines, we did not fit the spectrum and instead used the photometric data from \citet{2018ApJ...867...62W} to derive the parameters given in Table \ref{tab.specfits}.  We also retrieved photometry for this star from the Pan-STARRS1 $3\pi$ stack images photometric catalog \citep{2016arXiv161205560C}; these data prefer a slightly cooler temperature ($\teff\approx 10,000$ K) but lack the significant constraints provided by the $u$-band.

\begin{figure*}[tbp]
    \centering
    \includegraphics[clip, trim=0.5cm 5.5cm 1.5cm 13cm, width=\columnwidth] {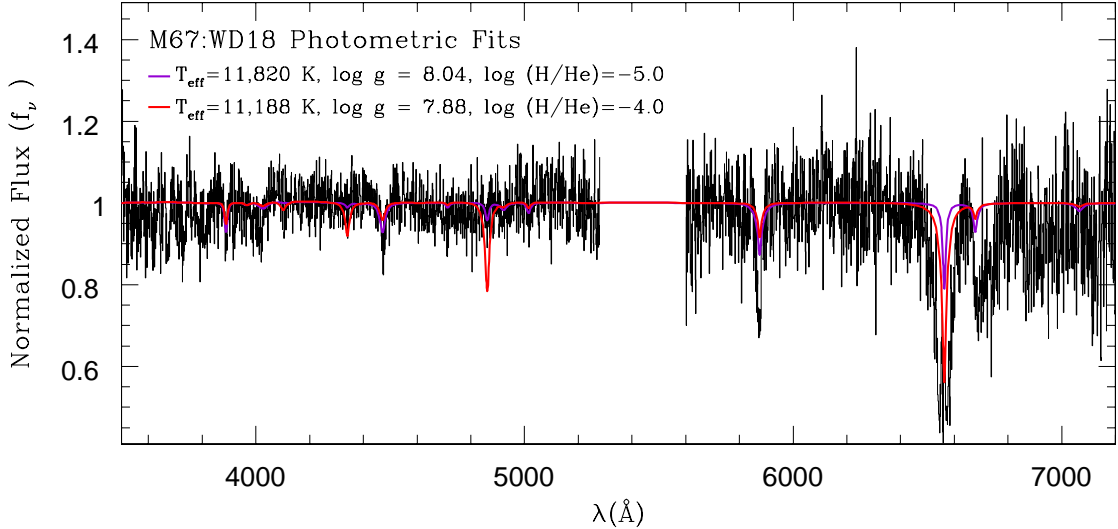}
    \caption{The normalized, observed spectrum of M67:WD18 (black) overlaid with normalized synthetic spectra for purely photometric WD parameter fits with assumed hydrogen abundances of $\log \mathrm{(H/He)} = -5.0$ (violet line with shallower H$\alpha$) and $\log \mathrm{(H/He)} = -4.0$ (red curve with deeper H$\alpha$).  Neither synthetic spectrum matches the observed spectral features well, and changes to \teff, \logg, and $\log \mathrm{(H/He)}$ cannot simultaneously match all significant spectral features.  We therefore suspect that this is a composite spectrum. }
    \label{fig.wd18_spec}
\end{figure*}

In Figure \ref{fig.wd18_spec}, we compare the observed spectrum to the synthetic spectra computed from the photometric solutions for assumed H abundances of $\log \mathrm{(H/He)} = -5$ and $\log \mathrm{(H/He)} = -4$.  Neither of these reproduce the observed spectrum well, and different assumed abundances are even poorer matches to the observed spectrum.   

Assuming that M67:WD18 is a single He-rich WD, the breadth of the H$\alpha$ line requires van der Waals broadening to explain; this broadening occurs in mixed atmospheres but at much lower \teff than indicated by the photometry.  A cooler, higher-pressure atmosphere would also explain the weakness or absence of the H$\beta$ absorption line.  However, this is completely at odds with the presence of \ion{He}{1} lines in the spectrum.

We therefore suspect that the spectrum of M67:WD18 is composite. Some spectral features, such as the absorption from $\sim 6650-6800$ \AA, look to be molecular bands from a cool K or M dwarf companion, while the width and depth of H$\alpha$ seem more indicative of a cool DA companion. The WD is unresolved and has no nearby neighbors in our deep MMT imaging \citep{2018ApJ...867...62W}. The parameter space is too unconstrained by the available data to (a) confirm that it is indeed a multiple star system, and (b) provide \teff and \logg for the individual components. As the H$\alpha$ feature likely does not originate from the He-rich WD, we simply adopt the photometric solution for a typical H abundance of $\log \mathrm{(H/He)} = -5$. We are reluctant to include this WD in an IFMR that assumes single star evolution, but its inclusion or exclusion does not affect our conclusions.

\subsubsection{Non-DA Parameters from the Literature}
Three of the WDs in our sample have well-determined parameters already published.  We now discuss each of these briefly in order to justify our adopted parameters for each WD.

\paragraph{Procyon B} Procyon B is a DQZ WD; its contribution to the IFMR was first discussed in detail by \citet{Liebert2013}, who used a spectroscopic mass determination of $M=0.553\pm 0.022 \msun$ and a total system age of $1.87\pm 0.13$ Gyr.  This resulted in a derived progenitor mass for Procyon B of $2.59_{-0.26}^{+0.44}\,\msun$.

Precision astrometry of the Procyon system by \citet{2015ApJ...813..106B} and \citet{2018RNAAS...2..147B} gives a significantly higher mass for Procyon B of $0.592\,\msun$.  In addition, asteroseismic analyses of Procyon A prefer older system ages.  \citet{2014ApJ...787..164G} derive an age of 2.4-2.8 Gyr using YREC stellar evolutionary models \citep{2008Ap&SS.316...31D}, while \citet{2019MNRAS.482..895S} derive a benchmark age range of $1.5-2.5$ Gyr from a variety of stellar models, though the majority of models they consider prefer ages of $2.0-2.5$ Gyr.

For this work, we adopt the dynamical mass for Procyon B of $M=0.592\pm 0.013\,\msun$ \citep{2018RNAAS...2..147B} and a spectroscopically-derived $\teff = 7740\pm 50$ K \citep{2002ApJ...568..324P}.   \citet{2019ApJ...885...74C} use a spectrophotometric fitting method to find a cooler $\teff=7585\pm 33$ K, though this results in a significantly lower mass ($0.554\pm 0.013\,\msun$) than the dynamical mass.  If we combine this more recent \teff determination with the dynamical mass, the model photometry would not agree with the known luminosity of Procyon B.  For that reason, we choose to adopt the \citet{2002ApJ...568..324P} \teff.  We note that if the \teff from \citet{2019ApJ...885...74C} is more precise, then the discrepancy of Procyon B with the DA IFMR would be even more severe; see Section \ref{sec.procyon_b}.

For the sake of consistency in our IFMR calculations, we adopt an asteroseismic age of $2.25\pm 0.25$ Gyr -- \citet{2019MNRAS.482..895S} derive an age of $2.2-2.3$ Gyr using PARSEC models, but we inflate the uncertainty to reflect the range in the majority of their models ($2.0-2.5$ Gyr).  Finally, we adopt a metallicity of [Fe/H]$=-0.05\pm 0.03$ \citep{2004AAp...413..251K}.

\paragraph{EGGR 316}  \object{EGGR 316} is a well-studied DBA in the Hyades open star cluster.  The recent spectral analysis by \citet{2018ApJ...857...56R} obtains $\teff=15,120 \pm 361$ K, $\logg=8.25\pm 0.09$, and $\log (\mathrm{H}/\mathrm{He}) = -4.68\pm 0.06$; these were derived using the same methodology as our spectral parameters in Section \ref{sec.params}.  For the Hyades, we adopt the PARSEC-derived cluster age of $700\pm25$ Myr \citep{2018ApJ...866...21C}.

\paragraph{DBH in Messier 47}  \citet{2019ApJ...880...75R} identify a massive magnetic helium-atmosphere WD (spectral type DBH) in the open star cluster Messier 47, \object[Gaia DR2 3029912407273360512]{Gaia 3029912407273360512}, which for simplicity we nickname M47:DBH.  We adopt their spectroscopic parameters, which were determined via the same methodology as our DB spectral parameters in Section \ref{sec.params}:  $\teff=32,100\pm 2700$ K and $M=1.06\pm 0.05\, \msun$.    For the cluster age and metallicity, we adopt the PARSEC-based determinations of \citet{2018AAp...616A..10G}: log age (yr) $=8.11^{+0.08}_{-0.06}$ and [Fe/H]$=+0.09$.

The adopted cluster parameters are summarized in Table \ref{tab.clparams}.

\begin{deluxetable}{lCCc}
\tablewidth{0pt}
\tablecaption{Adopted Cluster Parameters \label{tab.clparams}}
\tablehead{\colhead{Cluster} & \colhead{age (Myr)} & \colhead{[Fe/H]} & \colhead{Reference}}
\startdata
Hyades & 700\pm 25 & +0.15 & 1 \\
M47    & 129^{+26}_{-17} & +0.09 & 2 \\
Procyon & 2250\pm 250 & -0.05 & 3,4 \\
M67 & 3540\pm 150 & +0.02 & 5 \\
\enddata
\tablerefs{ (1) \citet{2018ApJ...866...21C} (2) \citet{2018AAp...616A..10G} (3) \citet{2019MNRAS.482..895S} (4) \citet{2004AAp...413..251K} (5) \citet{2015MNRAS.450.2500B} }
\end{deluxetable}

\section{Modeling Uncertainties in the IFMR \label{sec.errors}}

\begin{figure}
    \centering
    \includegraphics[width=0.9\columnwidth]{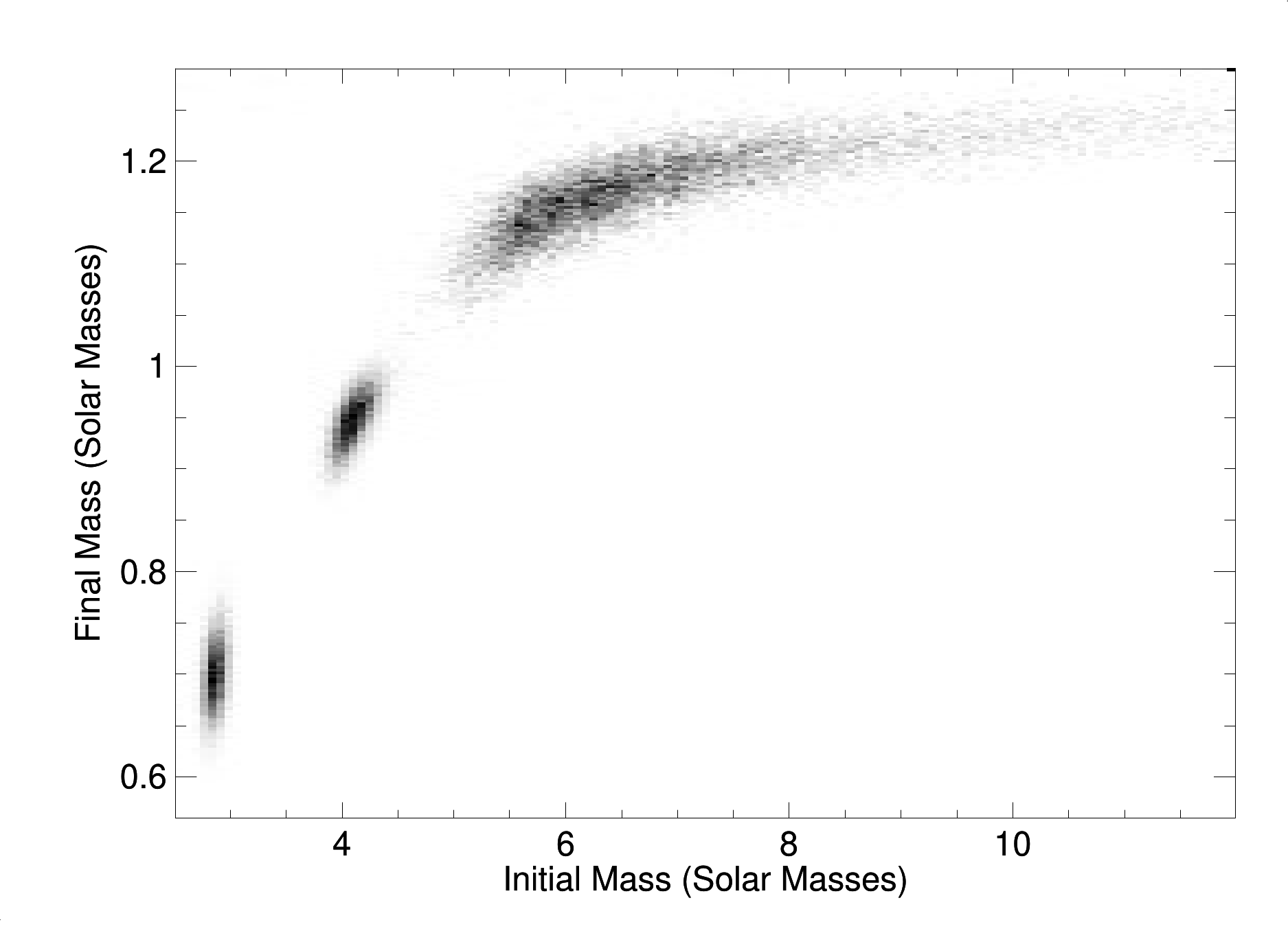}
    \caption{Examples of the correlated errors in the IFMR for three DA WDs with high signal-to-noise observations: \object[WD 0421+162]{WD $0421+162$} (bottom left), \object[Cl* NGC 3532 RK 1]{NGC 3532 RK 1} (center), and \object[WD 0605+241]{NGC 2168:LAWDS 2} (top right).  Grayscale indicates the density of points resulting from $10^4$ simulated observations of each WD using published errors in \teff and \logg, and assuming no uncertainty in the cluster age.  This figure originally appeared in  \citet{2020IAUS..357..179W}; reproduced with permission.}
    \label{fig.ifmr_sim_ex}
\end{figure}

Traditionally, the uncertainties in the location of WDs in the IFMR are portrayed by orthogonal error bars in $M_i$ and $M_f$ \citep[e.g.,][]{2009ApJ...693..355W,2018ApJ...866...21C}.  However, in many cases the uncertainties in these quantities are highly correlated. A higher-mass WD typically requires a larger $\tau_\mathrm{cool}$ to reach a given \teff, which for a given cluster age implies a shorter nuclear lifetime for the progenitor, which in turn implies a higher progenitor mass.  The slope of this correlation in the IFMR plane varies from WD to WD and is highly sensitive to the ratio of the uncertainty in $\tau_\mathrm{cool}$ to the derived progenitor nuclear lifetime \citep[Figure~\ref{fig.ifmr_sim_ex}, ][]{2020IAUS..357..179W}. 

\subsection{Model Errors for Individual White Dwarfs}

We therefore set out to determine the actual shape of the $M_i,\,M_f$ uncertainties for each of the non-DA WDs included in our semi-empirical IFMR via a Monte Carlo simulation.  The calculation consists of $10^6$ simulated measurements of each WD's \teff and \logg.  The values of each are drawn randomly assuming a Gaussian distribution centered on the best fit \teff and \logg with a standard deviation equal to the stated uncertainties on each parameter.  We assume the \teff and \logg are uncorrelated measurements, though in fact they are weakly correlated in many cases.  Each simulated \teff and \logg is then run through the same series of model-dependent calculations to derive $M_i$ and $M_f$.  

For each simulated observation, the cluster age is randomly chosen from a flat distribution centered on the adopted cluster age.  The best choice of age uncertainty distribution is not obvious \emph{a priori}, and many sources of open cluster ages do not specify the uncertainty distribution.  We tested four different error assumptions -- flat and Gaussian distributions in both linear and logartihmic distributions -- and the assumed distribution has no impact on our qualitative conclusions for this work.  

\begin{deluxetable}{lCCCC}
\tablewidth{0pt}
\tablecaption{PARSEC-derived WD Progenitor Masses\label{tab.minit}}
\tablehead{\colhead{WD} & \dcolhead{M_\mathrm{f}} &  \dcolhead{M_\mathrm{i}} & \dcolhead{dM_\mathrm{i,rand}} & \dcolhead{dM_\mathrm{i,sys}} \\
 & \msun & \msun & \msun & \msun}
\startdata
Hyades:EGGR 316 & 0.74\pm 0.06 & 3.26 & ^{+0.16}_{-0.12} & ^{+0.07}_{-0.07} \\
M47:DBH & 1.06\pm 0.05 & 6.73 & ^{+2.80}_{-1.10} & ^{+1.12}_{-1.02} \\
Procyon B & 0.59\pm 0.01 & 2.37 & ^{+0.03}_{-0.03} & ^{+0.26}_{-0.18} \\
M67:WD18 & 0.61\pm 0.14 & 1.58 & ^{+0.05}_{-0.03} & ^{+0.03}_{-0.03} \\
M67:WD21 & 0.53\pm 0.05 & 1.51 & ^{+0.01}_{-0.01} & ^{+0.02}_{-0.02} \\
M67:WD30 & 0.71\pm 0.09 & 1.52 & ^{+0.01}_{-0.01} & ^{+0.02}_{-0.02} \\
\enddata
\tablecomments{$M_\mathrm{f}$ is repeated from Table \ref{tab.specfits} for the reader's convenience.  Uncertainties are 68th percentile. Random errors include only propagated uncertainties in WD parameters; systematic errors include only cluster age uncertainty.} 
\end{deluxetable}

\begin{figure}
    \centering
    \includegraphics[width=0.9\columnwidth]{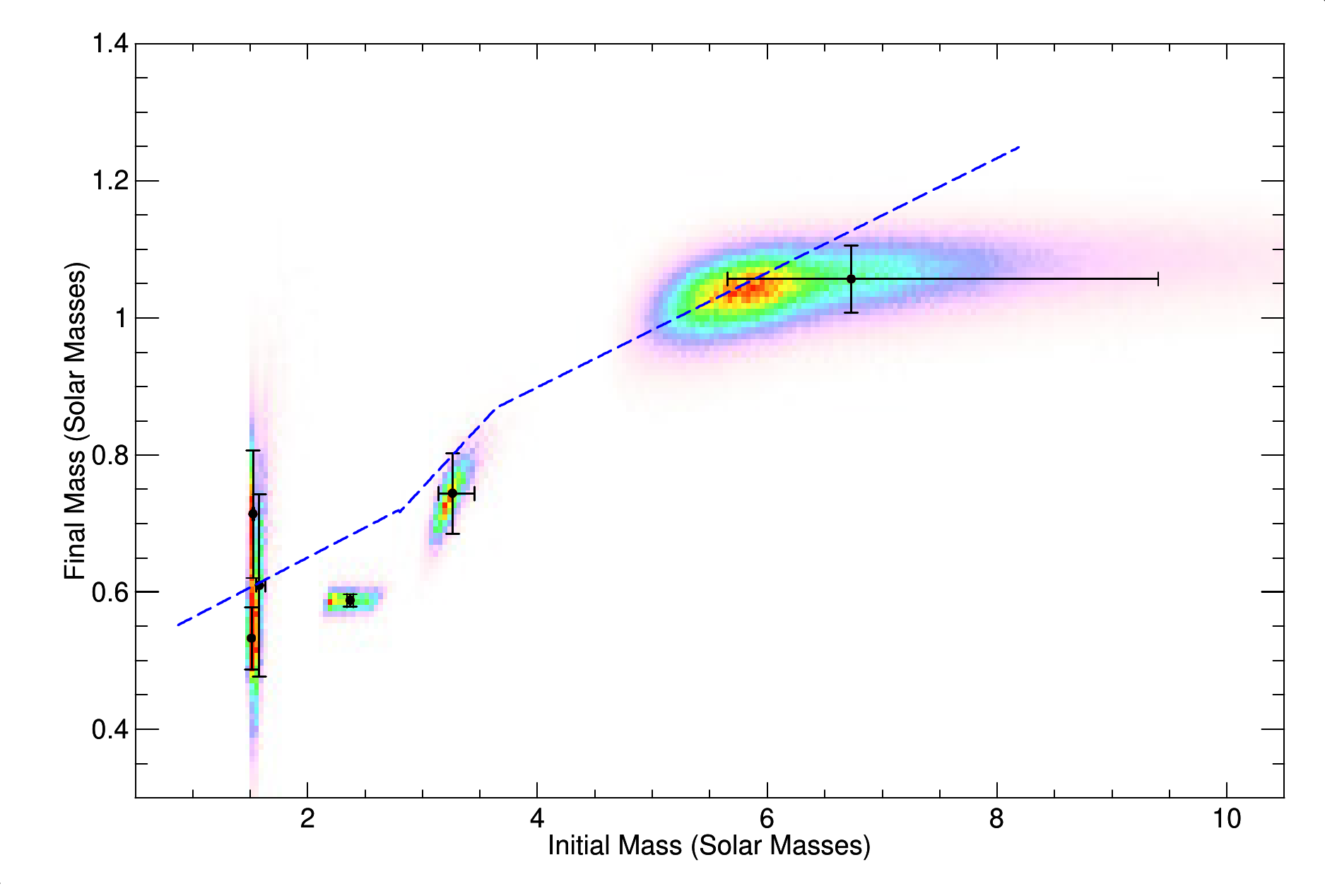}
    \caption{Comparison of the semi-empirical initial-final mass relation for non-DA white dwarfs (black points) to the piece-wise linear DA IFMR \citep[][dashed blue line]{2018ApJ...866...21C}.  Black points indicate the data from Table~\ref{tab.minit}, while the colored shading indicates the density of points in our simulation of IFMR uncertainties.  The density shading is rescaled for each WD such that the most common outcomes are shaded red.   The error bars indicate the range encompassed by 68\% of each WD's simulations.  In spite of large uncertainties, the non-DA WDs appear to lie below the DA IFMR.}
    \label{fig.ifmr}
\end{figure}

In Figure \ref{fig.ifmr}, we present a density plot of the simulated $M_i,\ M_f$ pairs for each WD. Error bars indicate the extent of 68\% of the simulated points centered on the median initial and final mass values for each WD; these median values are presented as the best-fit solutions in Table \ref{tab.minit}.  This analysis was accomplished by slightly modifying the IDL procedure density.pro \citep{densitypro}. 

Generally, the location of the $M_i,\,M_f$ point calculated from the best-fit WD parameters and assumed cluster age tend to lie at higher $M_i$ values than the highest density of simulated output points.  This raises a concern that the derived $M_i, M_f$ point may not accurately reflect the location of the true IFMR.  

Additional testing revealed that the apparent discrepancy between the calculated $M_i, M_f$ location and the simulation density distribution is almost exclusively due to the steep slope of the relation between progenitor masses and nuclear lifetimes; the location of the calculated $M_i, M_f$ point accurately reflects the median of both parameters in the simulations.  

\subsection{Modeling Uncertainty in the Shape and Location of the IFMR}

To be certain that the correlated errors and the offset between the median and mode of the simulated uncertainties does not bias the derivation of the semi-empirical IFMR, we conducted a further simulation of WD observations in order to compare an input IFMR with a recovered ``observed" IFMR.

Our simulation is designed to determine if the correlated errors result in a bias in the derived shape and slope of the IFMR.  We begin by generating stellar populations in 7 simulated cluster with known log ages $=7.75$ and $8.00-9.67$ in 0.33 dex increments.  Within each cluster, we draw $10^5$ stars with initial masses $\leq 8~\msun$ following a Salpeter IMF distribution and nuclear lifetimes shorter than the cluster age; nuclear lifetimes are determined from PARSEC v.~1.2S models with $Z=0.019$ \citep{2012MNRAS.427..127B,2014MNRAS.444.2525C,2015MNRAS.452.1068C}, retrieved using the CMD 3.4 interface\footnote{\href{http://stev.oapd.inaf.it/cgi-bin/cmd}{stev.oapd.inaf.it/cgi-bin/cmd}}.  Each input star is assumed to become a single WD with a mass given by the piecewise linear PARSEC IFMR of \citet{2018ApJ...866...21C}.  This input IFMR is derived using primarily DA WDs, so the thick H envelope, C/O core evolutionary models of \citet{2020ApJ...901...93B} are used to determine the \teff and \logg for each WD given the known cooling time (cluster age minus nuclear lifetime).  For now we ignore the complication that WDs resulting from single-star evolution with masses $\gtrsim 1.05~\msun$ likely have oxygen-neon cores, not C/O cores.

Our next step is to model the uncertainties in \teff and \logg resulting from spectral observations of each generated WD.  While in reality these uncertainties will vary depending on the details of observations, we chose to use typical uncertainties published in the recent literature for open cluster WDs: $\sigma_{\logg} = 0.05\,\mathrm{dex}$ and $\sigma_{\teff} = 0.025\times \teff$. The observed parameters, $T_\mathrm{eff,obs}$ and $\logg_\mathrm{obs}$ are then calculated by drawing the error in each value from a Gaussian distribution with a standard deviation of the adopted uncertainty in each parameter.   

Each simulated WD is then run through our IFMR machinery to determine its observed $M_i$ and $M_f$. Because spectroscopically derived parameters for cool DAs are affected by systematic issues in model atmospheres, and because these cool WDs are generally fainter than observational magnitude limits, we discard (without redrawing) any simulated WDs with $T_\mathrm{eff,obs} <  12,000$ K.  We also discard without redrawing all simulated WDs with derived cooling ages longer than the input cluster age, since the calculated nuclear lifetime of the progenitor would be negative.  

\begin{figure}
    \centering
    \includegraphics[width=0.9\columnwidth]{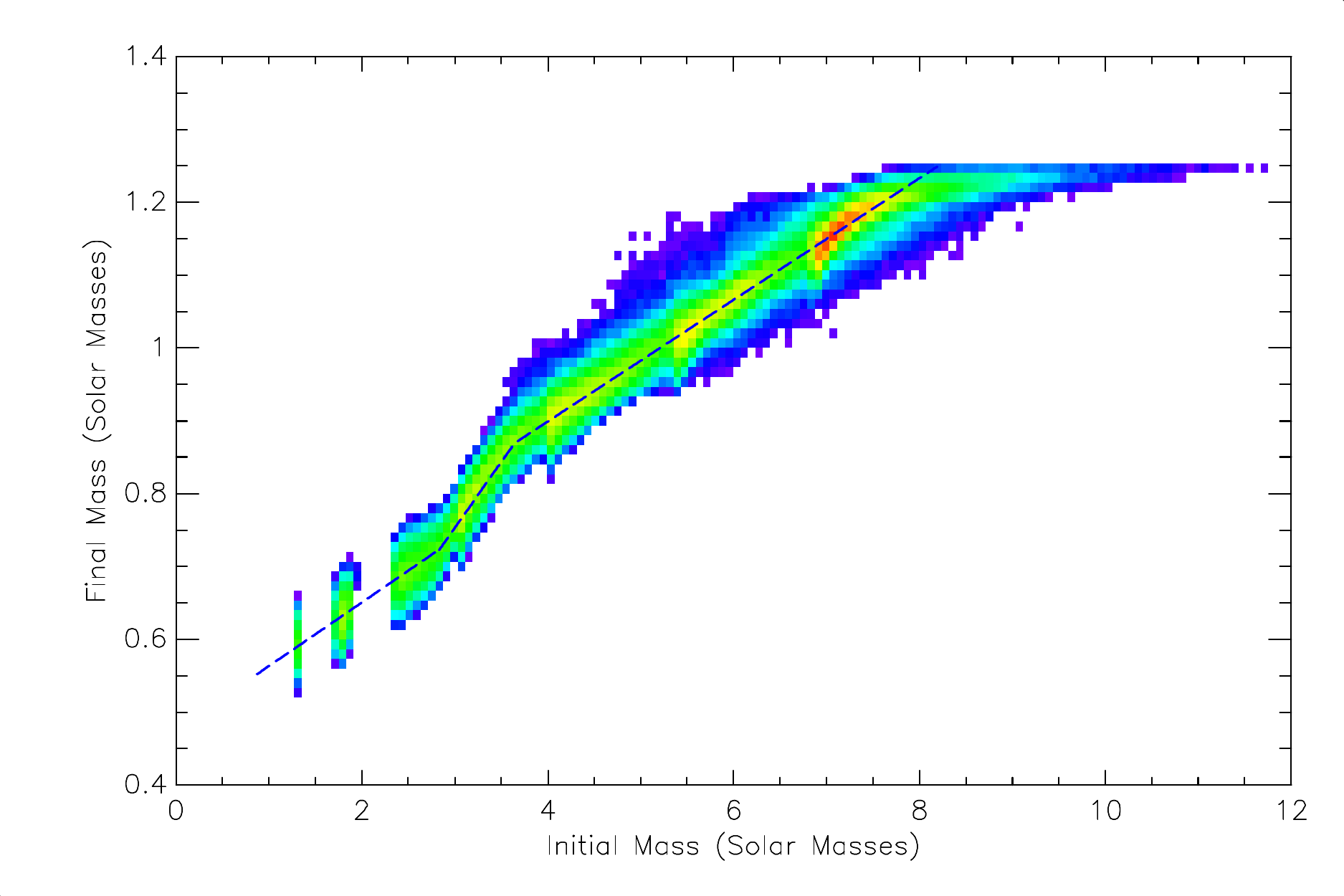}
    \caption{A density plot of the recovered semi-empirical IFMR compared with the input IFMR \citep[dashed line, from][]{2018ApJ...866...21C} from the Monte Carlo simulation described in the text. The density gradient is logarithmic.  Based on these results, we conclude that the correlated errors in the IFMR for individual WDs resulting from the uncertainties in \teff and \logg measurements such as those in Figure \ref{fig.ifmr_sim_ex} do not translate to significant biases in the shape and location of the ensemble semi-empirical IFMR, at least when uncertainties in assumed cluster ages are ignored.}
    \label{fig.sim_ifmr_obs}
\end{figure}

The recovered IFMR are shown in Figure \ref{fig.sim_ifmr_obs} as a logarithmic density plot compared to the input IFMR, indicated by the dashed line.  We see that the recovered IFMR is in excellent qualitative agreement with the input IFMR, though with a slight asymmetric tail to higher recovered initial masses at the high $M_i$ end. The lowest-density regions (blue and purple shades) of the recovered IFMR indicate a small number of simulations per pixel and are associated with  simulated WDs with $\tcool > \tnuc$.

Based on these simulations, we conclude that the correlated uncertainties in initial and final masses for individual WDs introduce no significant biases in shape and location of the semi-empirical IFMR recovered from an ensemble of WDs, at least when only uncertainties in the observations of \teff and \logg are considered.  Uncertainties in the ages of the input clusters will introduce systematic errors for all WDs in a given cluster, and that effect has not been considered here.

\section{The Non-DA Initial-Final Mass Relation: Discussion \label{sec.db_ifmr}}

With a modicum of confidence in our understanding of the uncertainties, we can now compare the position of non-DA WDs in the initial-final mass plane to the semi-empirical IFMR derived from (primarily) DA WDs.   We exclude M67:WD27 and M67:WD31 from the IFMR analysis, as the WD masses derived from their low \logg values are sufficiently low to indicate that either these two DCs are the He-core results of binary evolution, or that they are not bona fide cluster members.

In Figure \ref{fig.ifmr}, we compare the non-DA WDs from this analysis to the piecewise-continuous linear fit derived from \citet{2018ApJ...866...21C}.  For both sets of data, we use the PARSEC stellar evolutionary models for deriving both star cluster ages and the progenitor masses as a function of nuclear lifetimes.  We find that the DBs are consistent with the DA-derived IFMR, but that Procyon B, a DQZ, has a mass 0.1 \msun lower than the DA IFMR.

\subsection{The outlier Procyon B \label{sec.procyon_b}}
This apparently low mass for Procyon B has been noted before \citep[e.g.,][]{Liebert2013}, and potential explanations are discussed thoroughly in \citet{2015ApJ...813..106B}.  In summary, the parameters of Procyon A and Procyon B as determined from precision astrometric measurements and astroseismic observations of Procyon A are consistent and well constrained.  It therefore is not possible to reconcile the location of Procyon B in the initial-final mass plane with the DA IFMR from measurement errors alone.  

\citet{2015ApJ...813..106B} suggest two solutions to the Procyon problem. First, perhaps the Procyon system underwent significant binary interactions in the past.
 In summary, the separation of the binary components prior to mass loss from Procyon B's progenitor star would have been $\sim 5$ AU.  This is large enough that the pair would not have shared a common envelope, but tidal interactions and mass transfer from the post-main sequence wind of Procyon B may have occurred.  If the total mass transfer were significant, the evolution of Procyon A would be sped up, making the system appear substantially younger than it actually is.

The second suggestion of \citet{2015ApJ...813..106B} is that the apparently low mass of Procyon B is simply due to large intrinsic scatter in the IFMR, i.e., there is some spread in the WD mass resulting from a progenitor star of a given mass.  This suggestion is based on the apparent scatter in the semi-empirical IFMR of \citet{2005MNRAS.361.1131F}.  

We are skeptical of this latter hypothesis. \citet{2015ApJ...807...90C} illustrate that this apparent scatter is significantly reduced when WD parameters and cluster ages are determined in an internally consistent manner, which was not the case in \citet{2005MNRAS.361.1131F}.   Further, \citet{2018ApJ...867...62W} show that the measured standard deviation of WD masses in M67 is $\approx 0.04 \msun$ once low-mass He core WDs and abnormally massive WDs potentially resulting from blue stragglers are excluded from the sample; this scatter is consistent with the observational uncertainties and constrains any intrinsic scatter in the M67 IFMR to be less than 0.02 \msun. 

Procyon B is significantly cooler than the other non-DA WDs analyzed here.  It also has a spectral type of DQZ in contrast to the DAs and other non-DAs in the IFMR; the models used to derive Procyon B's temperature and cooling age could be systematically off. Indeed, \citet{2019ApJ...885...74C} argue that there appears to be a systematic shift in the parameters derived for DQ WDs, potentially due to problems with UV carbon opacities in the model atmosphere.  Other systematic effects such as gravitational settling of $^{22}$Ne, not included in the evolutionary models of \citet{2020ApJ...901...93B}, could also be affecting the derived cooling age \citep[e.g.,][]{2008ApJ...677..473G}.

Yet even if Procyon B's cooling age is treated as unknown, its initial mass is constrained by the age of Procyon A (assuming no significant binary interactions) to be $\gtrsim 1.5\,\msun$, and its WD mass would still be below the DA IFMR, though at a lower significance.  For these reasons, neither intrinsic scatter nor observational scatter in the IFMR seem likely explanations for Procyon B.

\subsection{Differences in the DA and non-DA IFMR?}
\begin{figure}
    \centering
    \includegraphics[width=0.9\columnwidth]{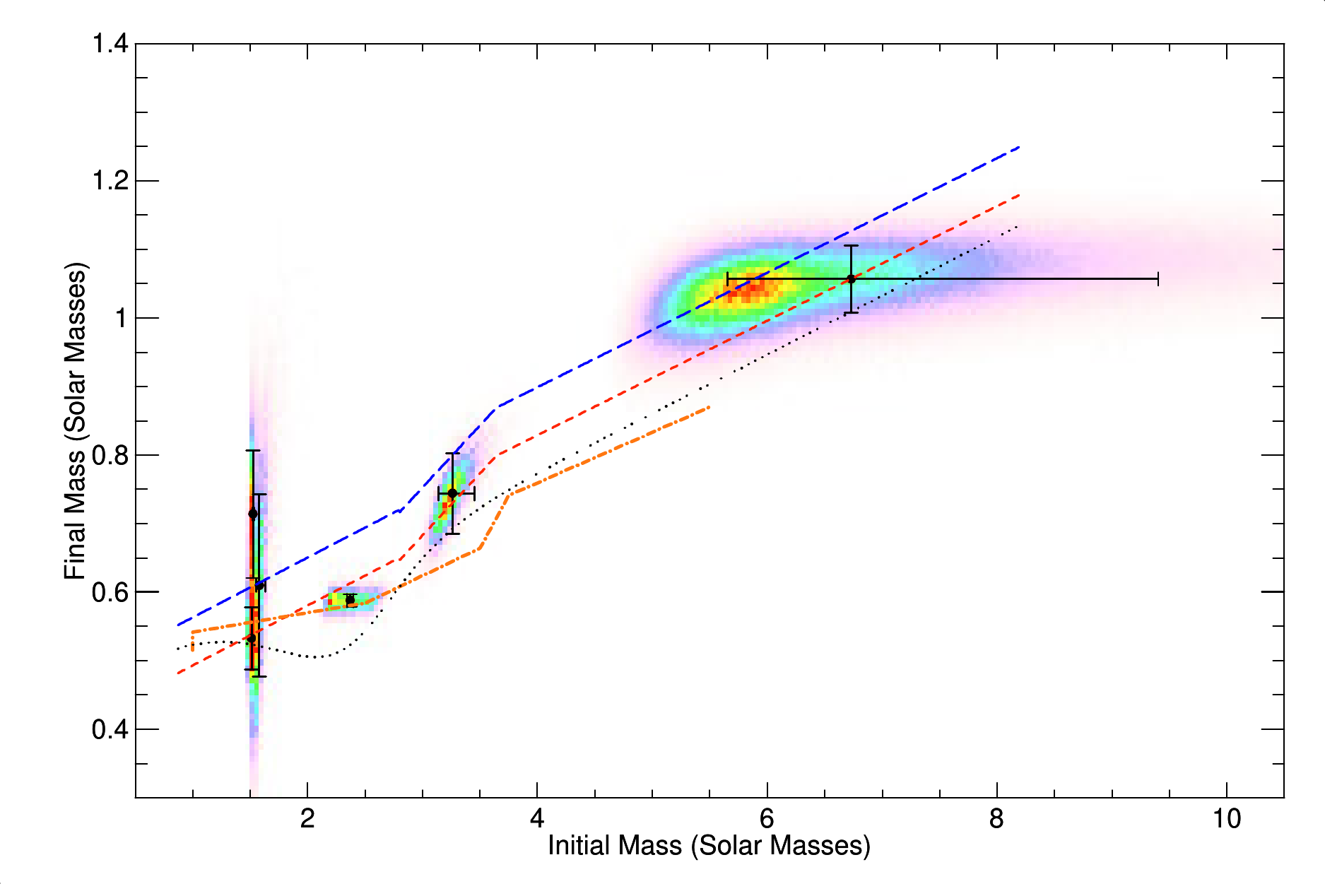}
    \caption{The same non-DA IFMR as in Figure \ref{fig.ifmr} compared to various IFMRs: the \citet{2018ApJ...866...21C} semi-empirical IFMR for DAs (blue, long-dashed line), the \citet{2018ApJ...866...21C} IFMR shifted to lower final masses by 0.07\,\msun (red, short-dashed line), the modelled non-DA IFMR of \citet{2009ApJ...704.1605A} (orange dash-dotted line), and the core mass at first thermal pulse relation from \citet{2013MNRAS.434..488M} (black dotted curve).   With the exception of M67:WD30, all the non-DA WDs in this sample including Procyon B are most consistent with the shifted DA IFMR, and reasonably consistent with the model-based IFMRs.}
    \label{fig.ifmr_shift}
\end{figure}

A third possible explanation for the position of Procyon B compared to the semi-empirical IFMR is that the non-DA IFMR differs from the DA IFMR. Figure \ref{fig.ifmr_shift} compares the semi-empirical non-DA IFMR and uncertainties with the DA IFMR of \citet{2018ApJ...866...21C}, the model-derived non-DA IFMR of \citet{2009ApJ...704.1605A}, the functional fit to model-based core masses of stars at the first thermal pulse from \citet{2013MNRAS.434..488M}, and the DA IFMR arbitrarily shifted by 0.07~\msun to lower WD masses.  With the exception of M67:WD30,  all of the non-DA WDs appear most consistent with the shifted IFMR and are consistent within errors of the two model-based curves.  

If the 0.07 \msun offset we observe is indeed physical in nature, an obvious potential explanation would be enhanced mass loss during the (V)LTP and the rejuvenated star's additional stint on the post-AGB evolutionary track.  Since the maximum mass of a WD's helium layer is $\approx 10^{-2}\,M_\mathrm{WD}$, much of the required additional 0.07 \msun of mass loss would have to come from the carbon-oxygen core of the star. This mass loss could be observable due to enhanced dust formation.  However, the observed mass lost following these late thermal pulses ($\sim 10^{-4}$) is more than two orders of magnitude lower than the observed difference we suggest might exist.  \citet{2005A&A...435..631A} report their models show an additional 0.013~\msun of mass loss during the hot post-AGB evolution following a VLTP, but even this mass loss would be nearly an order of magnitude too small. For this reason and additional considerations outlined below, we have serious doubts about the physical reality of a significant difference between the DA and non-DA IFMRs.

Even if the mass loss during the (V)LTP and renewed post-AGB evolution of the star is enhanced, or even if some other physical mechanism during stellar evolution affects the mass of non-DAs, the difference in WD masses hinted at by the non-DA IFMR would be in conflict with multiple measurements of the mean masses of DA and non-DA WDs, which typically are observed to be the same within 0.01 or 0.02 \msun.  

The possibility exists that our observations and/or analysis techniques have introduced a systematic error in our derived initial or final masses. As described in Section \ref{sec.notes}, all three M67 non-DAs included in the IFMR have some oddity in either the spectral fits or the resulting parameters.  We have double-checked the data and reductions from these stars and others observed on the same night with the same setup, and we do not see evidence for any obvious systematic error.

As discussed by \citet{Salaris2009} and illustrated clearly by \citet{2018ApJ...866...21C}, a precision IFMR should be constructed from self-consistent models and analysis.  The non-DA IFMR we present is not produced in a fully consistent manner.  Our WD parameters come from a mixture of photometric and spectroscopic fits, except for Procyon B, which has an astrometrically measured mass.  Our adopted star cluster ages are generally derived from isochrone fitting using PARSEC isochrones, but the Procyon system age was derived from a combination of asteroseismological measurements with stellar evolutionary models, though for the sake of consistency we prefer the age derived using PARSEC models.  

As mentioned previously, the spectral models for non-DA WDs contain complex physics that is not fully understood and are currently undergoing cycles of revision, and potentially this could bias our results.  For example, we have not applied 3D corrections to WD spectroscopic parameters proposed by \citet{2021MNRAS.501.5274C}, as their corrections are likely to change as additional and improved microphysical calculations are included in model atmospheres.  Yet these same systematics should also affect the mass determinations of non-DA field WDs -- and the mean masses and dispersions of field DA and DB WDs are not significantly different \citep[e.g.,][]{2019ApJ...871..169G}.

Conservatively, our analysis may be biased by overconfidence in our use of Procyon B in the non-DA IFMR.  If we ignore Procyon B, then the significance of any difference between the DA and non-DA semi-empirical IFMRs is greatly reduced and could be explained away as being due to small number statistics.  As described above, there are reasons to be suspicious of the inclusion of Procyon B in an IFMR analysis, since we are not certain that its past interactions with Procyon A were negligible.  

Additional non-DA WDs are clearly needed to be studied and added to the non-DA semi-empirical IFMR if we are to confirm or refute the potential differences between the IFMRs.  Because of the relative rarity of non-DA WDs and because of the exquisite observations required for spectral fitting, progress is likely to be slow if we limit ourselves to analysis of high signal-to-noise spectral observations of open cluster non-DA WDs.  
Two alternative observational strategies may permit more efficient progress to be made on the non-DA IFMR.   First, common proper motion binaries with a non-DA WD component may be more numerous than open cluster non-DA WDs once Gaia results are thoroughly searched.  The IFMR constructed from these binaries tends to have larger uncertainty in $M_i$ due to more uncertain ages of the binary companion, but sample size can overcome this drawback.  

Second, for non-DA WDs there has been substantial progress in obtaining precision WD \teff and \logg from multi-band photometric modeling, especially if near UV photometry is available outside of the Rayleigh-Jeans tail of the WD spectral energy distribution.  Combined with astrometric confirmation that a given WD is a member of a cluster or a wide binary, these photometric models may reduce the need for very high signal-to-noise optical spectroscopy \citep{2019ApJ...871..169G,2019ApJ...882..106G}.  However, some spectroscopy is likely to remain necessary in order to confirm the non-DA nature of candidate WDs and to provide reasonable limits on the H abundance in the WD atmosphere.

\subsection{Conclusions\label{sec.conclusions}}

In this paper, we have presented results from a study of the semi-empirical IFMR for non-DA WDs.  As our discussions touched on several subjects, we summarize our main conclusions here.

\begin{itemize}
\item We present a sample of non-DA WDs that are known astrometric members of open star clusters or, in the case of Procyon B, a component in a well-studied binary system.  For those WDs without previously published spectral parameters, we present spectroscopic and photometric fits to synthetic atmospheres in order to determine \teff, \logg, and associated uncertainties in each parameter. Using DB evolutionary models to determine \tcool for each WD in the sample, we construct a semi-empirical IFMR for non-DA WDs.

\item Because the uncertainties in the initial-final mass plane are highly correlated, we conducted Monte Carlo simulations of both the true uncertainties in the parameters derived for individual WDs and in the shape and location of the ensemble semi-empirical IFMR.  We find that the correlated errors for individual WDs are non-Gaussian and non-symmetric, but when a large sample of WDs is considered, the inferred semi-empirical IFMR is not greatly biased in spite of the correlated errors in $M_f$ and $M_i$. This conclusion holds true for both DA and non-DA WDs.

\item The non-DA IFMR is mildly inconsistent with published semi-empirical DA IFMRs; the non-DA WDs appear to be roughly 0.07 \msun less massive for a given initial mass than DA WDs.  If physical, this offset could be due to unexpected enhanced mass loss from the WDs during the thermal pulse and renewed post-AGB evolution of the star. This offset is significantly larger than models and observations of mass loss during LTPs and VLTPs, and it is inconsistent with the observed agreement of the mean masses of DA and non-DA field WDs.  We are therefore skeptical that the apparent difference in the DA and non-DA IFMRs is physically real.

\item The apparent differences in the IFMRs may be biased by the inclusion of Procyon B in the sample.  Procyon B has been previously noted as being inconsistent with the semi-empirical IFMR.  Potential past binary interactions between Procyon A \& B may explain its discrepant point in the IFMR.

\item A significantly larger sample size of non-DA WDs is needed to determine whether or not the non-DA IFMR differs from that of the DA IFMR.  If there is indeed a significant difference, this would imply part of our understanding of the progenitors of non-DA WDs is incomplete.

\end{itemize}

\begin{acknowledgements}
The authors wish to recognize and acknowledge the very significant cultural role and reverence that the summit of Maunakea has always had within the indigenous Hawaiian community.  We are most fortunate to have the opportunity to conduct observations from this mountain.  

This work has been supported by the National Science Foundation under grants awards AST-1910551 and AST-0602288, and supported by Cottrell College Science Award 22706 from the Research Corporation for Science Advancement.  A.B. acknowledges support from the Natural Sciences and Engineering Research Council (NSERC) of Canada through an Alexander Graham Bell Graduate Scholarship.

J.Barnett is grateful for the service of M.~Wood and K.~Montgomery on his M.S. thesis committee and for their helpful comments and suggestions during this project. The authors also thank Pierre Bergeron for his assistance in setting up this collaboration. The authors also thank the anonymous referee for her/his efforts and suggestions. The exquisite observing skills of K.H.R.~Rubin were crucial to the success of this project.   
\end{acknowledgements}

\facility{Keck I (LRIS-B)}

\software{IRAF \citep{1986SPIE..627..733T,1993ASPC...52..173T}}

\bibliographystyle{aasjournal}

\end{document}